\PassOptionsToPackage{unicode}{hyperref}
\PassOptionsToPackage{hyphens}{url}
\PassOptionsToPackage{dvipsnames,svgnames,x11names}{xcolor}
\documentclass[
]{article}
\usepackage{amsmath,amssymb}
\usepackage{lmodern}
\usepackage{iftex}
\ifPDFTeX
  \usepackage[T1]{fontenc}
  \usepackage[utf8]{inputenc}
  \usepackage{textcomp} 
\else 
  \usepackage{unicode-math}
  \defaultfontfeatures{Scale=MatchLowercase}
  \defaultfontfeatures[\rmfamily]{Ligatures=TeX,Scale=1}
\fi
\IfFileExists{upquote.sty}{\usepackage{upquote}}{}
\IfFileExists{microtype.sty}{
  \usepackage[]{microtype}
  \UseMicrotypeSet[protrusion]{basicmath} 
}{}
\makeatletter
\@ifundefined{KOMAClassName}{
  \IfFileExists{parskip.sty}{%
    \usepackage{parskip}
  }{
    \setlength{\parindent}{0pt}
    \setlength{\parskip}{6pt plus 2pt minus 1pt}}
}{
  \KOMAoptions{parskip=half}}
\makeatother
\usepackage{xcolor}
\usepackage{color}
\usepackage{fancyvrb}

\DefineVerbatimEnvironment{Highlighting}{Verbatim}{commandchars=\\\{\}}
\newenvironment{Shaded}{}{}

\newcommand{\CommentTok}[1]{\textcolor[rgb]{0.38,0.63,0.69}{\textit{#1}}}

\newcommand{\NormalTok}[1]{#1}
\newcommand{\OperatorTok}[1]{\textcolor[rgb]{0.40,0.40,0.40}{#1}}

\usepackage{graphicx}
\makeatletter
\def\maxwidth{\ifdim\Gin@nat@width>\linewidth\linewidth\else\Gin@nat@width\fi}
\def\maxheight{\ifdim\Gin@nat@height>\textheight\textheight\else\Gin@nat@height\fi}
\makeatother
\setkeys{Gin}{width=\maxwidth,height=\maxheight,keepaspectratio}
\makeatletter
\def\fps@figure{htbp}
\makeatother
\makeatletter
\@ifundefined{pandocbounded}{%
  \newsavebox\pandoc@box
  \newcommand*\pandocbounded[1]{
    \sbox\pandoc@box{#1}%
    \Gscale@div\@tempa{\textheight}{\dimexpr\ht\pandoc@box+\dp\pandoc@box\relax}%
    \Gscale@div\@tempb{\linewidth}{\wd\pandoc@box}%
    \ifdim\@tempb\p@<\@tempa\p@\let\@tempa\@tempb\fi
    \ifdim\@tempa\p@<\p@\scalebox{\@tempa}{\usebox\pandoc@box}%
    \else\usebox\pandoc@box%
    \fi%
  }%
}{}
\makeatother
\providecommand{\tightlist}{%
  \setlength{\itemsep}{0pt}\setlength{\parskip}{0pt}}
\NewDocumentCommand\citeproctext{}{}
\NewDocumentCommand\citeproc{mm}{%
  \begingroup\def\citeproctext{#2}\cite{#1}\endgroup}
\makeatletter
 \let\@cite@ofmt\@firstofone
 \def\@biblabel#1{}
 \def\@cite#1#2{{#1\if@tempswa , #2\fi}}
\makeatother
\newlength{\cslhangindent}
\newlength{\csllabelwidth}
\newenvironment{CSLReferences}[2] 
 {\begin{list}{}{%
  \setlength{\itemindent}{0pt}
  \setlength{\leftmargin}{0pt}
  \setlength{\parsep}{0pt}
  \ifodd #1
   \setlength{\leftmargin}{\cslhangindent}
   \setlength{\itemindent}{-1\cslhangindent}
  \fi
  \setlength{\itemsep}{#2\baselineskip}}}
 {\end{list}}
\usepackage{calc}

\ifLuaTeX
\usepackage[bidi=basic]{babel}
\else
\usepackage[bidi=default]{babel}
\fi
\babelprovide[main,import]{american}

\def\languageshorthands#1{}
\ifLuaTeX
  \usepackage{selnolig}  
\fi
\IfFileExists{bookmark.sty}{\usepackage{bookmark}}{\usepackage{hyperref}}
\IfFileExists{xurl.sty}{\usepackage{xurl}}{} 
\hypersetup{
  pdftitle={halox: Dark matter halo properties and large-scale structure
calculations using JAX},
  pdfauthor={Florian Kéruzoré, Lance A. Moreau},
  pdflang={en-US},
  colorlinks=true,
  linkcolor={Maroon},
  filecolor={Maroon},
  citecolor={Blue},
  urlcolor={Blue},
  pdfcreator={LaTeX via pandoc}}

\title{halox: Dark matter halo properties and large-scale structure
calculations using JAX}

\definecolor{c53baa1}{RGB}{83,186,161}
\definecolor{c202826}{RGB}{32,40,38}

\usepackage[affil-it]{authblk}
\usepackage{orcidlink}
\author[1%
  *%
  ]{Florian Kéruzoré%
    \,\orcidlink{0000-0002-9605-5588}\,%
    }
\author[3,2,1%
  *%
  ]{Lance A. Moreau%
    \,\orcidlink{0009-0004-5742-8478}\,%
    }

\affil[1]{High Energy Physics Division, Argonne National Laboratory,
Lemont, IL 60439, USA%
  }
\affil[2]{Department of Astronomy, University of Maryland, College Park,
MD 20742-2421, USA%
  }
\affil[3]{Department of Physics, University of Maryland, College Park,
MD 20742-2421, USA%
  }
\affil[*]{These authors contributed equally.}
\date{29 April 2026}

\begin{document}
\maketitle

\section{Summary}\label{summary}

Dark matter halos are fundamental structures in cosmology, forming the
gravitational potential wells hosting galaxies and clusters of galaxies.
Their properties and statistical distribution (including the halo mass
function) are invaluable tools to infer the fundamental properties of
the Universe. The \texttt{halox} package is a JAX-powered Python library
enabling differentiable and accelerated computations of key properties
of dark matter halos, and of the halo mass function. The automatic
differentiation capabilities of \texttt{halox} enable its usage in
gradient-based workflows, \emph{e.g.}. in efficient Hamiltonian Monte
Carlo sampling or machine learning applications. The acceleration
capabilities of \texttt{halox} enable significant speedups over existing
packages such as \texttt{colossus} on GPU architectures, while offering
comparable performance on CPUs.

\section{Statement of need}\label{statement-of-need}

In cosmology and astrophysics, modeling dark matter halos is central to
understanding the large-scale structure of the Universe and its
formation. Recently, the AI-driven advent of novel computational
frameworks such as JAX (\citeproc{ref-Bradbury:2018}{Bradbury et al.,
2018}) has led to the development of differentiable and
hardware-accelerated software to simulate and model physical processes,
with \emph{e.g.}. Brax (\citeproc{ref-Brax:2021}{Freeman et al., 2021})
and JAX, MD (\citeproc{ref-Jaxmd:2020}{Schoenholz \& Cubuk, 2020}). The
increasing complexity of cosmological data and astrophysical models has
motivated the wide adoption of this framework in cosmology, where
JAX-powered software has been published to address a wide variety of
scientific goals, including modeling fundamental cosmological
quantities, with, \emph{e.g.}., JAX-cosmo
(\citeproc{ref-Campagne:2023}{Campagne et al., 2023}) and LINX
(\citeproc{ref-Giovanetti:2024}{Giovanetti et al., 2024}); simulating
density fields and observables, with, \emph{e.g.}., SHAMNet
(\citeproc{ref-Hearin:2022}{Hearin et al., 2022}), DISCO-DJ
(\citeproc{ref-Hahn:2024}{Hahn et al., 2024}), JAXpm
(\citeproc{ref-Jaxpm:2025}{Differentiable Universe Initiative, 2025}),
and JAX-GalSim (\citeproc{ref-JaxGalSim:2025}{JAX-GalSim developpers,
2025}; \citeproc{ref-Mendoza:2025}{Mendoza et al., 2025}); emulating
likelihoods for accelerated inference, with, \emph{e.g.}.,
CosmoPower-JAX (\citeproc{ref-Piras:2023}{Piras \& Spurio Mancini,
2023}) and candl (\citeproc{ref-Balkenhol:2024}{Balkenhol et al.,
2024}); or modeling various physical properties of dark matter halos,
such as mass accretion history (Diffmah,
\citeproc{ref-Hearin:2021}{Hearin et al., 2021}), galaxy star formation
history (Diffstar, \citeproc{ref-Alarcon:2023}{Alarcon et al., 2023}),
halo concentration (Diffprof,
\citeproc{ref-Stevanovich:2023}{Stevanovich et al., 2023}), gas-halo
connection (picasso, \citeproc{ref-Keruzore:2024}{Kéruzoré et al.,
2024}), and halo mass function (\citeproc{ref-Buisman:2025}{Buisman et
al., 2025})\footnote{Note that halox also provides an implementation of
  the halo mass function, but chooses a lighter, halo model-based
  approach; see \textbf{Software design}.}.

The \texttt{halox} library offers a JAX implementation of some widely
used properties of dark matter halos. The use of JAX as a backend allows
these functions to be compiled and GPU-accelerated, enabling
high-performance computations; and automatically differentiable,
enabling their efficient use in gradient-based workflows, such as
sensitivity analyses, Hamiltonian Monte-Carlo sampling for Bayesian
inference, or machine learning-based methods. In addition, expensive
computations of large-scale structure properties are further accelerated
using neural network emulators, preserving hardware acceleration and
differentiability while enabling faster calculations thanks to
approximate calculations (see the \textbf{Software design} section).

\section{State of the field}\label{state-of-the-field}

Many toolkits focused on halo modeling have been developed, such as,
\emph{e.g.}., halofit (\citeproc{ref-Smith:2003}{Smith et al., 2003}),
halotools (\citeproc{ref-Hearin:2017}{Hearin et al., 2017}), colossus
(\citeproc{ref-Diemer:2018}{Diemer, 2018}), and pyCCL
(\citeproc{ref-Chisari:2019}{Chisari et al., 2019}). While JAX-powered
software exists for a wide variety of cosmological applications (see
\textbf{Statement of need}), the properties implemented in
\texttt{halox} (listed in \textbf{Software design}), although available
in other libraries focused on halo modeling, do not currently have
publicly available, differentiable and hardware-accelerated
implementations.

\section{Software design}\label{software-design}

\subsection{Available physical
quantities}\label{available-physical-quantities}

The \texttt{halox} library seeks to provide JAX-based implementations of
common models of dark matter halo properties and of large-scale
structure. At the time of writing (software version 2.1.1), this
includes the following properties:

\begin{itemize}
\tightlist
\item
  Cosmological quantities: \texttt{halox} relies on JAX-cosmo
  (\citeproc{ref-Campagne:2023}{Campagne et al., 2023}) for
  cosmology-dependent calculations, and includes wrapper functions to
  compute some additional properties, such as critical density
  \(\rho_{\rm c}\) and differential comoving volume element
  \({\rm d}V_{c} / {\rm d}\Omega {\rm d}z\).
\item
  Radially-dependent physical properties of NFW and Einasto dark matter
  halos. Our NFW and Einasto implementations are based on the analytical
  derivations of Łokas \& Mamon (\citeproc{ref-Lokas:2001}{2001}) and
  Retana-Montenegro et al. (\citeproc{ref-Retana-Montenegro:2012}{2012})
  respectively, and include the following quantities:

  \begin{itemize}
  \tightlist
  \item
    Matter density \(\rho(r)\);
  \item
    Enclosed mass \(M(\leq r)\);
  \item
    Gravitational potential \(\phi(r)\);
  \item
    Circular velocity \(v_{\rm circ}(r)\);
  \item
    Velocity dispersion \(\sigma_{v}(r)\) (NFW only);
  \item
    Projected surface density \(\Sigma(r)\) (NFW only).
  \end{itemize}
\item
  Concentration-mass relations: There are implementations of several
  relations including:

  \begin{itemize}
  \tightlist
  \item
    Duffy et al. (\citeproc{ref-Duffy:2008}{2008})
  \item
    Klypin et al. (\citeproc{ref-Klypin:2011}{2011})
  \item
    Prada et al. (\citeproc{ref-Prada:2012}{2012})
  \item
    Child et al. (\citeproc{ref-Child:2018}{2018}) (for all halo and
    relaxed halo populations)
  \end{itemize}
\item
  Large-scale structure: Building upon the power spectra computations
  implemented in JAX-cosmo, \texttt{halox} provides implementations of
  the RMS variance of the matter distribution in spheres of radius
  \(R\), \(\sigma(R)\). It also includes a wrapper function to perform
  the computation within the Lagrangian radius of a halo of mass \(M\),
  \(\sigma(M)\).
\item
  Halo mass function (HMF): The HMF model of Tinker et al.
  (\citeproc{ref-Tinker:2008}{2008}), predicting
  \({\rm d}N / {\rm d} \ln M\) as a function of halo mass \(M\),
  redshift \(z\), and cosmology.
\item
  Halo bias: The linear bias model of Tinker et al.
  (\citeproc{ref-Tinker:2010}{2010}) as a function of halo mass \(M\),
  redshift \(z\), and cosmology.
\item
  Overdensities: All properties in \texttt{halox} can be computed for
  spherical overdensity (SO) halo masses defined for any critical
  overdensity value. Convenience functions are provided to convert halo
  properties from one critical overdensity to another, or to convert
  critical overdensities to and from mean matter overdensities.
\end{itemize}

\subsection{Automatic differentiation and hardware
acceleration}\label{automatic-differentiation-and-hardware-acceleration}

All calculations available in \texttt{halox} are written using JAX and
JAX-cosmo. As a result, all functions can be compiled just-in-time using
\texttt{jax.jit}, hardware-accelerated, and are automatically
differentiable with respect to their input parameters, including halo
mass, redshift, and cosmological parameters. In addition, all JAX
transformations can be used on \texttt{halox} functions, including
native vectorization and parallelization using \emph{e.g.}.
\texttt{jax.vmap}.

\subsection{Emulation}\label{emulation}

\(\sigma^2(R)\) is the variance of the fluctuations of the matter
density field in a sphere of radius \(R\). At redshift \(z\) and for
cosmological parameters \(\Omega\), it is given by:

\[\sigma^2(R,z,\Omega) = \frac{1}{2 \pi^2} \int_0^\infty k^2 W^2(k, R) P(k, z, \Omega) {\rm d}k,\]

where \(k\) denotes spatial frequency, \(P(k,z,\Omega)\) is the power
spectrum, and \(W\) is the Fourier transform of the spherical top-hat
window function. \(\sigma\) is an essential ingredient in computing both
halo mass function and halo bias in most standard parameterizations
(\emph{e.g.}., \citeproc{ref-Tinker:2010}{Tinker et al., 2010}), and the
numerical integration is computationally expensive, and often the
primary bottleneck in such calculations.

To tackle this issue, \texttt{halox} also includes an emulated
calculation of \(\sigma\), as a function of mass (the Lagrangian mass
contained in a radius \(R\)), redshift, and cosmological parameters. Our
emulator consists of a multi-layer perceptron with three hidden layers,
each of width 64. The emulator was trained on the halox \(\sigma(M)\)
implementation. The training set is taken from a Sobol sample over
log(M), log(1+z), and the cosmological parameters \(\Omega_b\),
\(\Omega_c\), \(h\), \(n_s\), and \(\sigma_8\).

\begin{figure}
\centering
\includegraphics[width=0.9\linewidth,height=\textheight,keepaspectratio]{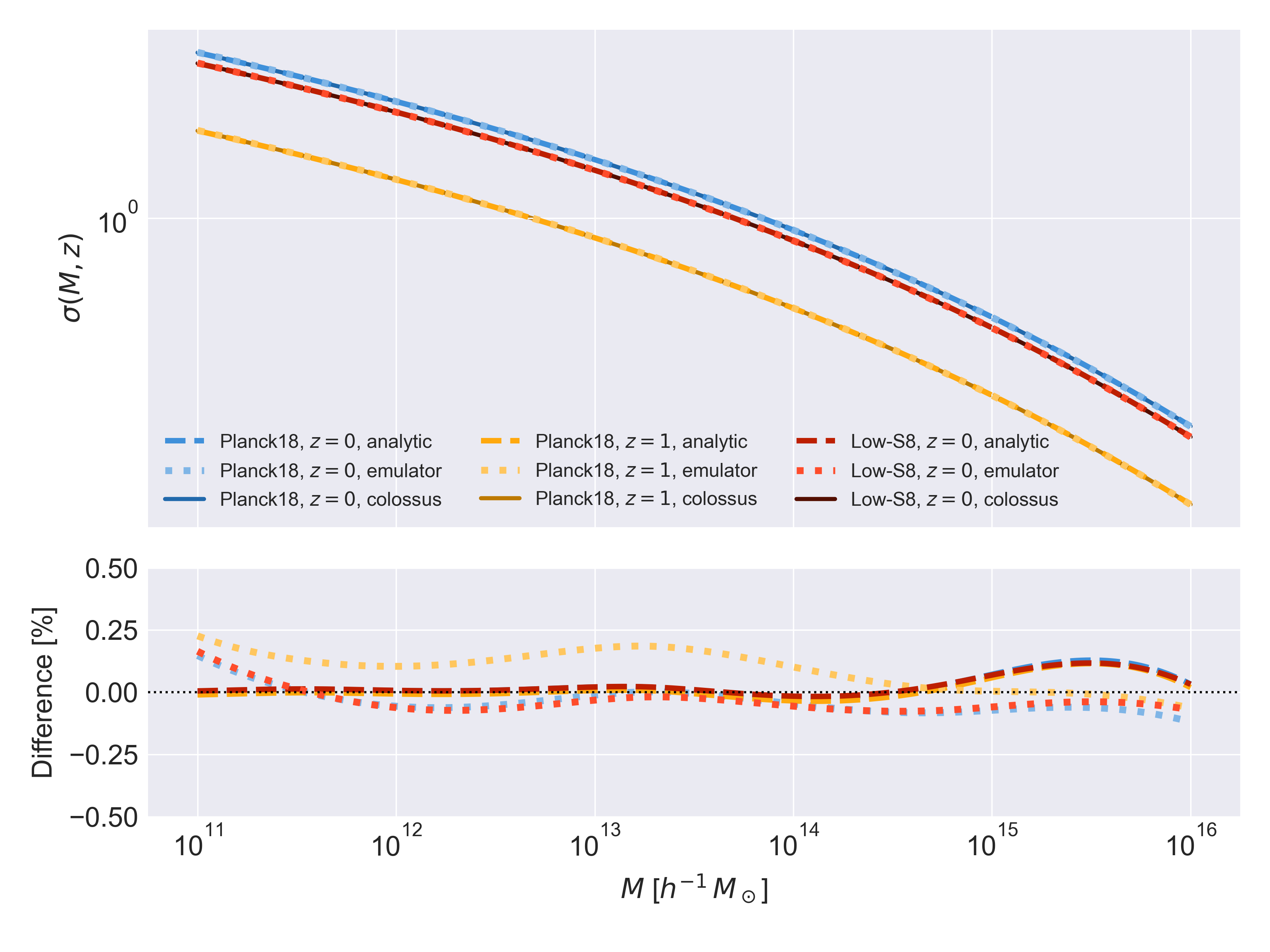}
\caption{\emph{Top:} \(\sigma(M)\) predicted using \texttt{colossus},
\texttt{halox}, and the \texttt{halox} emulator, varying redshift and
cosmology. \emph{Bottom:} Fractional difference between \texttt{halox}
predictions and \texttt{colossus}.}\label{fig:figure1}
\end{figure}

\begin{figure}
\centering
\includegraphics[width=0.9\linewidth,height=\textheight,keepaspectratio]{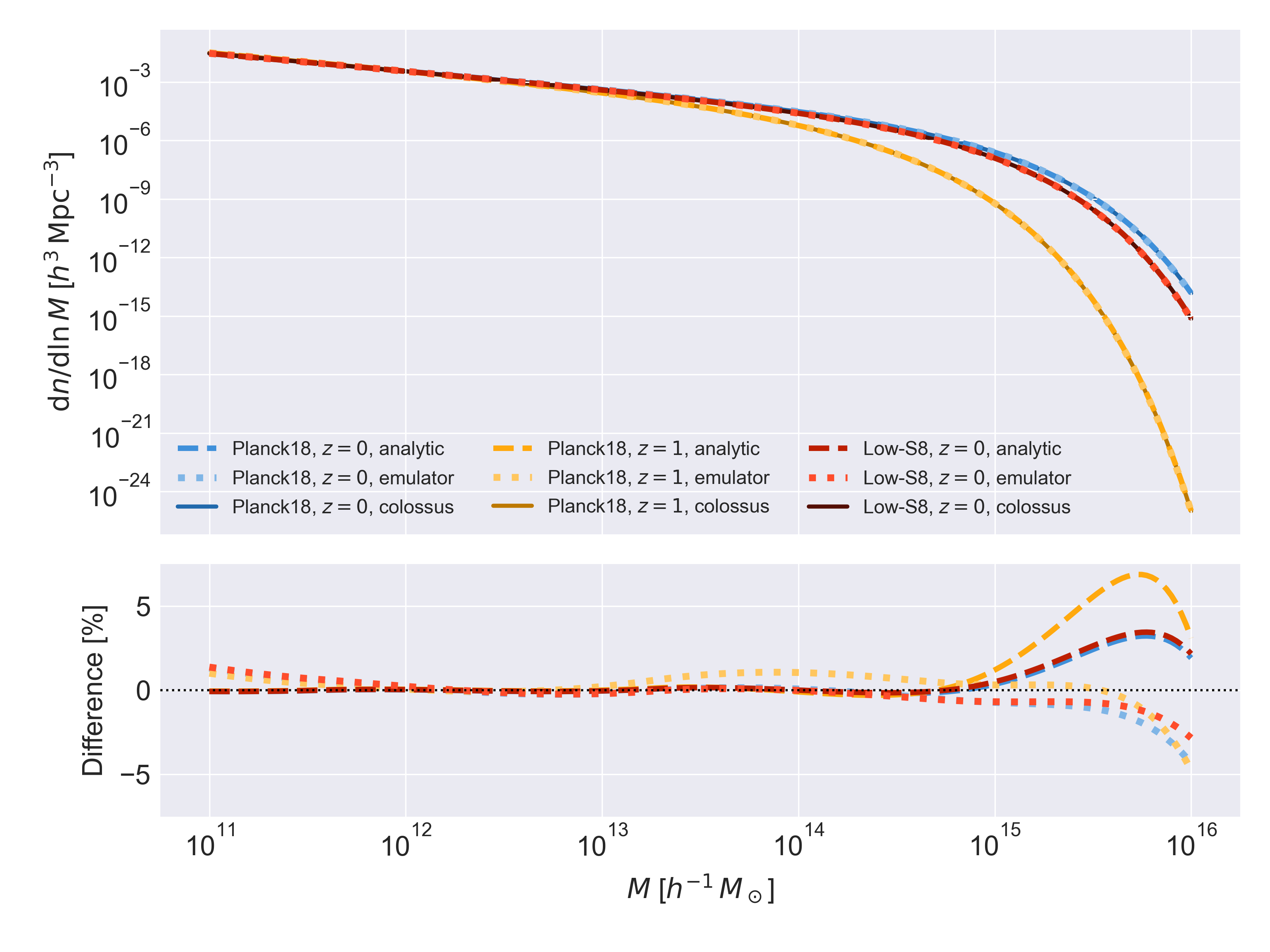}
\caption{Same as Figure 1 for the halo mass
function.}\label{fig:figure2}
\end{figure}

The emulator is accurate to within a percent for both \(\sigma(M)\) and
the halo bias, and within a few percent for the HMF across the tested
parameter space. \autoref{fig:figure1} and \autoref{fig:figure2} show
the accuracy of emulator-based predictions of \(\sigma(M)\) and of the
halo mass function for different cosmologies up to \(z=1\),
demonstrating remarkable accuracy, with an increase of error in the
regime of extremely massive halos
(\(M_{200c} > 10^{15} \, h^{-1} M_\odot\)), which are very rare
occurences in both simulations and observations.

To compute \(\sigma(M)\), the HMF, or the halo bias using the emulator,
users may simply instantiate the emulator, then pass it in as an
optional argument to the original \(\sigma(M)\) function in halox:

\begin{Shaded}
\begin{Highlighting}[]
\NormalTok{emu }\OperatorTok{=}\NormalTok{ emus.sigmaM.SigmaMEmulation()               }\CommentTok{\# instantiate emulator}
\NormalTok{sigma\_a }\OperatorTok{=}\NormalTok{ halox.lss.sigmaM(M, z, cosmo)           }\CommentTok{\# analytic sigma(M)}
\NormalTok{sigma\_e }\OperatorTok{=}\NormalTok{ halox.lss.sigmaM(M, z, cosmo, emu}\OperatorTok{=}\NormalTok{emu)  }\CommentTok{\# emulated sigma(M)}
\end{Highlighting}
\end{Shaded}

The same calling sequence can be used to compute halo mass function or
halo bias using the \(\sigma(M)\) emulator.

\section{Research impact statement}\label{research-impact-statement}

At the time of writing, \texttt{halox} is being used in several upcoming
analyses, including the development of a differentiable generator of
halo catalogs and the creation of a suite of realistic simulations of
the extragalactic millimeter-wave sky. In addition, it has already been
referenced in cosmology publications, \emph{e.g.}., LSST Dark Energy
Science Collaboration et al. (\citeproc{ref-Desc:2026}{2026}), Alarcon
et al. (\citeproc{ref-Alarcon:2026}{2025}).

\subsection{Performance benchmarks}\label{performance-benchmarks}

\begin{figure}
\centering
\pandocbounded{\includegraphics[keepaspectratio]{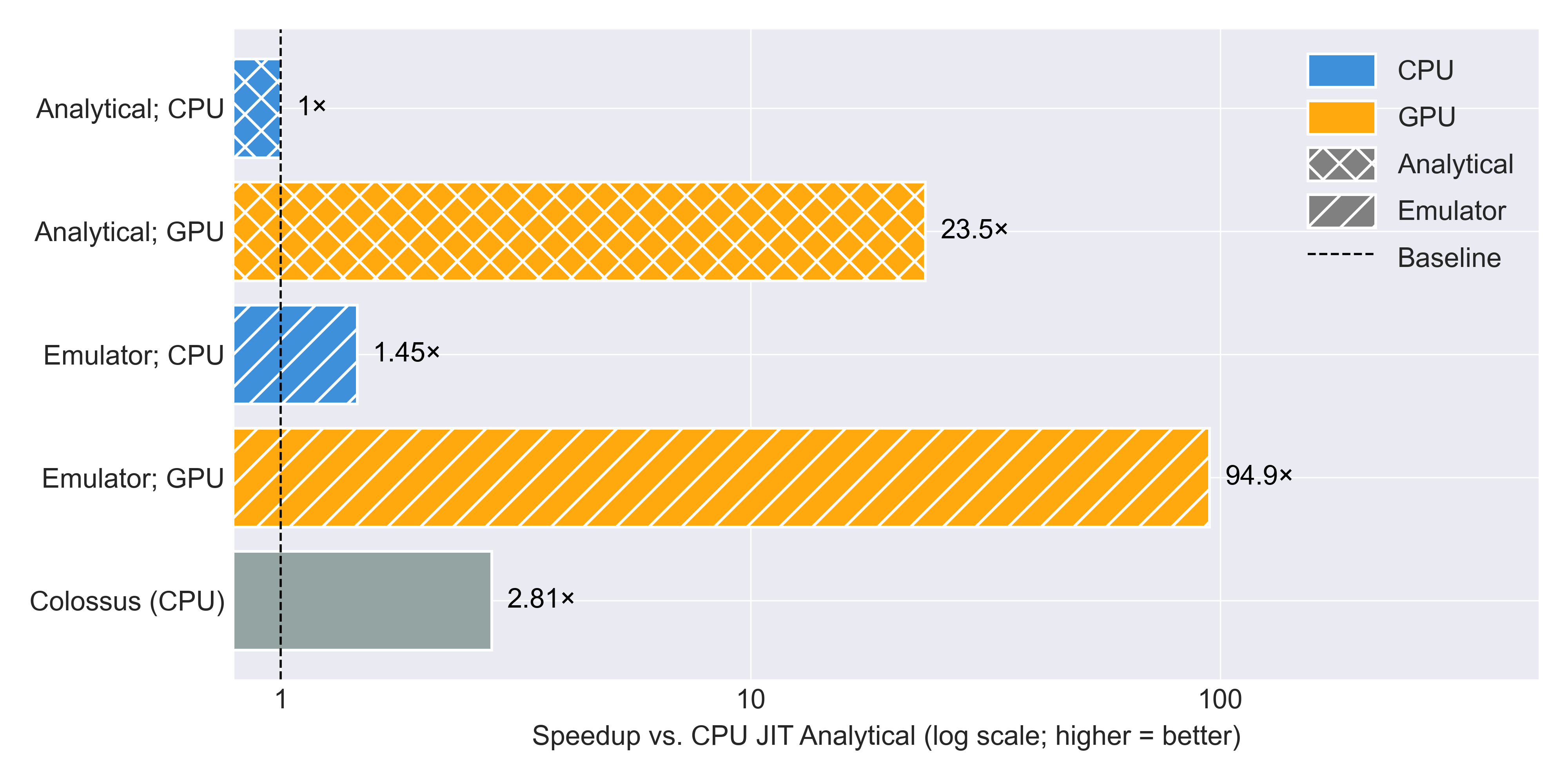}}
\caption{The performance of HMF computation for the halox package on
different architectures and against \texttt{colossus}. All CPU
executions are still slower than \texttt{colossus} irrespective of
emulation. GPU architecture enables further speedup, allowing for faster
computations than \texttt{colossus} both with and without emulation,
with significant speedup when using the emulated function over the
standard calculation. \label{fig:figure3}}
\end{figure}

\autoref{fig:figure3} shows the performance of halo mass function
computations in \texttt{halox} across hardware configurations,
benchmarked against our slowest evaluation (analytical computation on
CPU). For comparison, we also compare to the performance of
\texttt{colossus} on the same computation\footnote{Our benchmarks were
  run on an AMD EPYC 7742 CPU (2.25GHz) and an NVIDIA A100-SXM4-40GB
  GPU, evaluating the HMF on a grid of 256 halo masses \(\times\) 256
  redshifts, for a fixed cosmology. All \texttt{halox} computations were
  run after just-in-time compilation.}. Three results stand out. First,
we see that \texttt{colossus} outperforms \texttt{halox} on CPU, by
about a factor of three for the analytic computation, owing to calls to
highly efficient libraries such as CAMB (\citeproc{ref-Lewis:2011}{Lewis
\& Challinor, 2011}) and well-optimized interpolation and integration
schemes. Second, using a GPU significantly accelerates \texttt{halox}
predictions, by a factor of over 20 for the analytic computation, and of
about 65 for the emulated version. Third, using the neural network
emulator as a backend for the \(\sigma(M)\) computation enables a
substantial speedup, up to \(95\times\) compared to baseline, and
\(34\times\) compared to \texttt{colossus}. These results demonstrate
the strong potential of \texttt{halox} in GPU-based cosmological
analyses, delivering considerable speedup in addition to automatic
differentiation. We also note that all computations were made at double
(FP64) precision; \texttt{halox} can be further accelerated, in
particular on GPUs, by dropping to single (FP32) precision.

\subsection{Validation}\label{validation}

All functions available in \texttt{halox} are validated against
existing, non-JAX-based software. Cosmology calculations are validated
against Astropy (\citeproc{ref-Astropy:2022}{Astropy Collaboration et
al., 2022}) for varying cosmological parameters and redshifts. Other
quantities are validated against either \texttt{colossus}
(\citeproc{ref-Diemer:2018}{Diemer, 2018}) or Gala
(\citeproc{ref-Gala:2017}{Price-Whelan, 2017}) for varying halo masses,
redshifts, critical overdensities, and cosmological parameters. These
tests are included in an automatic CI/CD pipeline on the GitHub
repository, and presented graphically in the online documentation.

\section{AI usage disclosure}\label{ai-usage-disclosure}

We acknowledge the use of Anthropic's Claude (Sonnet 4.5, Opus 4.5, Opus
4.6) in the development of \texttt{halox} (code and documentation). We
note that no AI was used in writing the unit test suite enforcing
accuracy of the software, which allowed us to ensure that the generated
code produced correct results.

\section{Acknowledgments}\label{acknowledgments}

We would like to thank Lindsey Bleem, Matt Becker, Jean-Éric Campagne,
Andrew Hearin, and Georgios Zacharegkas for useful discussions and
feedback on \texttt{halox} and on this manuscript. This work was
supported in part by the U.S. Department of Energy, Office of Science,
Office of Workforce Development for Teachers and Scientists (WDTS) under
the Science Undergraduate Laboratory Internships (SULI) Program. Argonne
National Laboratory's work was supported by the U.S. Department of
Energy, Office of Science, Office of High Energy Physics, under contract
DE-AC02-06CH11357.

\section*{References}\label{references}
\addcontentsline{toc}{section}{References}

\protect\phantomsection\label{refs}

\end{document}